\documentclass{IEEEtran}
\usepackage{cite}
\usepackage[pdftex]{graphicx}
\usepackage[cmex10]{amsmath}
\usepackage{amssymb}
\usepackage{url}
\usepackage{multirow}

\title{Informed Group-Sparse Representation for Singing Voice Separation}

\author{Tak-Shing~T.~Chan,~\IEEEmembership{Member,~IEEE} and Yi-Hsuan~Yang,~\IEEEmembership{Member,~IEEE}%
\thanks{Manuscript received Month xx, 2016; revised Month xx, 2016; accepted Month xx, 2017. Date of publication Month xx, 2017; date of current version Month xx, 2016. The associate editor coordinating the review of this manuscript and approving it for publication was Prof.~xxxxxxxx xxxxxxxx.}%
\thanks{The authors are with the Research Center for Information Technology Innovation, Academia Sinica, Taipei 11564, Taiwan (e-mail: takshingchan@citi.sinica.edu.tw; yang@citi.sinica.edu.tw).}%
\thanks{Digital Object Identifier 10.1109/LSP.2017.2647810}}

\IEEEpubid{\copyright~2017 IEEE. Personal use is permitted. For any other purposes, permission must be obtained from the IEEE by emailing pubs-permissions@ieee.org.}

\begin{document}

\maketitle

\begin{abstract}
Singing voice separation attempts to separate the vocal and instrumental parts of a music recording, which is a fundamental problem in music information retrieval. Recent work on singing voice separation has shown that the low-rank representation and informed separation approaches are both able to improve separation quality. However, low-rank optimizations are computationally inefficient due to the use of singular value decompositions. Therefore, in this paper, we propose a new linear-time algorithm called informed group-sparse representation, and use it to separate the vocals from music using pitch annotations as side information. Experimental results on the iKala dataset confirm the efficacy of our approach, suggesting that the music accompaniment follows a group-sparse structure given a pre-trained instrumental dictionary. We also show how our work can be easily extended to accommodate multiple dictionaries using the DSD100 dataset.
\end{abstract}


\section{Introduction}
\label{sec:intro}

\IEEEPARstart{T}{he} problem of recovering unknown sources from observed mixtures, known as source separation, has been successfully applied to various fields including communications, medical imaging, and audio \cite{Comon10}. Such an inverse problem is well-posed \cite{Tikhonov95,Tarantola05} if it has a unique solution that depends on the data continuously. More specifically, in singing voice separation (SVS) \cite{Vembu05,Ryynanen08,Huang12,Sprechmann12}, the aim is to separate the singing voice from the instrumentals, which has numerous applications in music information retrieval \cite{Tsai12,Nakano14}. Unfortunately, in the case of one microphone and more than one sources, a unique solution is mathematically impossible, so monaural source separation is generally ill-posed \cite{Vincent14}. The prevalent approach to an ill-posed problem, say $(\arg)\min_X F(X)$, is to formulate a regularizer $R(X)$ \cite{Tikhonov95,Cappe96,Kowalski08} which incorporates some prior assumptions:
\begin{equation}
\label{eq:1}
\min_X F(X)+\lambda R(X),
\end{equation}
where $\lambda$ is a regularization parameter to be determined empirically (e.g., by cross validation). Usually, $R(X)$ is chosen to favor a particular class of solutions. For example, one of the most popular regularizers is a sparsifier:
\begin{equation}
\label{eq:2}
\min_X F(X)+\lambda\left\|X\right\|_1,
\end{equation}
where the elementwise $\ell_1$-norm $\left\|X\right\|_1$ encourages the matrix to be sparse \cite{Tibshirani96}. This regularizer dominates an area of research known as sparse coding \cite{Foldiak90,Olshausen97}, which is frequently used in audio separation \cite{Virtanen03,Plumbley10}. Another attractive regularizer is a low-rank regularizer:
\begin{equation}
\label{eq:3}
\min_X F(X)+\lambda\left\|X\right\|_*,
\end{equation}
where the trace norm $\left\|X\right\|_*$ or the sum of the singular values of $X$ is employed to favor low-rank solutions \cite{Fazel01,Srebro05}. This regularizer is often seen in singing voice separation in recent years \cite{Huang12,Yang13,Ikemiya15,Chan15}. Last but not least, in informed audio source separation \cite{Bryan13,Liutkus13,Chan15}, we want to fuse external annotations into the final optimized solution, which is most helpful if the annotations are close to the correct solution (as in score-informed separation \cite{Ewert14}). This requirement can be met by the following regularizer \cite{Chen13,Lefevre14,Jeong15}:
\begin{equation}
\label{eq:4}
\min_X F(X)+\frac{\lambda}{2}\left\|X-X_0\right\|_F^2,
\end{equation}
where $X$ can be a magnitude spectrogram and $X_0$ denotes the annotations on it. Such annotations may be obtained from the corresponding musical scores or from specific techniques for tracking the vocal melody contour. As evidenced above, regularization is quite versatile and it can incorporate both model assumptions and model answers into the problem itself. Still more information can be packed into the regularizer through a dictionary, as we will see in the related work below.

\IEEEpubidadjcol

\subsection{Related Work}

The robust principal component analysis (RPCA) decomposes an input matrix $X$ into a low-rank matrix $A$ and a sparse matrix $E$ \cite{Lin09,Candes11}:
\begin{equation}
\label{eq:5}
\min_{A,E}\left\|A\right\|_*+\lambda\left\|E\right\|_1~\text{s.t.}~X=A+E.
\end{equation}
Unlike traditional PCA, RPCA is robust against gross errors. For music spectrograms, if we assume that the instrumentals are repetitive \cite{Rafii13} and the vocals are sparse \cite{Huang12}, then the RPCA can be applied to the SVS problem \cite{Huang12}. This assumption is reasonable because musical instruments tend to have relatively stable and regular harmonic patterns while we can only sing one note at a time. The main drawback to this approach is that the resulting sparse matrix often contains instrumental solo or percussion \cite{Su13,Yang12}. A partial solution for this problem is to incorporate reliable annotations for the sparse part using informed RPCA (hereafter RPCAi) \cite{Chen13}:
\begin{equation}
\label{eq:6}
\min_{A,E}\left\|A\right\|_*+\lambda\left\|E\right\|_1+\frac{\gamma}{2}\left\|E-E_0\right\|_F^2~\text{s.t.}~X=A+E,
\end{equation}
where $E_0$ denotes the annotations (e.g., the pointwise product of $X$ and a binary matrix). Sometimes $A$ is not itself low-rank but is instead low-rank in a given dictionary. In this case, the low-rank representation (LRR) can be used \cite{Liu13,Yang13}:
\begin{equation}
\label{eq:7}
\min_{Z,E}\left\|Z\right\|_{*} +\lambda\left\|E\right\|_1~\text{s.t.}~X=DZ+E,
\end{equation}
where $D$ is a predefined (or pre-learned) dictionary such that $A=DZ$. It can be seen that LRR is an extension of RPCA because \eqref{eq:7} reduces to \eqref{eq:5} when $D=I$.

While we can perform dictionary-informed separation by simply combining LRR with the informed-separation-norm \eqref{eq:4}, LRR uses the singular value decomposition (SVD), an $O{\left(n^3\right)}$ algorithm, which can be slow for larger datasets. In light of this, we will propose a dictionary-based group-sparse representation (GSR) model for SVS informed by annotated melodies\footnote{Melody annotations have long been a feature in SVS datasets such as MIR-1K \cite{Hsu10} and iKala \cite{Chan15}. If unavailable, approximations can still be obtained using existing pitch tracking algorithms, such as MELODIA \cite{Salamon12}.} (GSRi). Our contributions are summarized in context in Table~\ref{tab:1}.

In what follows, we present our informed group-sparse representation model in Section~\ref{sec:gsri} and the experimental results in Section~\ref{sec:exp}. The extension to multiple dictionaries is also described and tested before we conclude in Section~\ref{sec:conc}.

\begin{table}
\renewcommand{\arraystretch}{1.2}
\caption{Our contributions in context (see Section~\ref{sec:intro}): $X$ is the input magnitude spectrogram, $A$ or $DZ$ are the resulting instrumentals, and $E$ is resulting vocals. Here $E_0$ denotes the vocal annotations and $D$ denotes an instrumental dictionary.}
\label{tab:1}
\centering
\begin{tabular}{|c|c|c|}
\hline
Method & Objective & Constraint\\
\hline\hline
RPCA \cite{Candes11} & $\left\|A\right\|_*+\lambda\left\|E\right\|_1$ & $X=A+E$\\
\hline
RPCAi \cite{Chen13} & $\left\|A\right\|_*+\lambda\left\|E\right\|_1+\frac{\gamma}{2}\left\|E-E_0\right\|_F^2$ & $X=A+E$\\
\hline
LRR \cite{Liu13} & $\left\|Z\right\|_*+\lambda\left\|E\right\|_1$ & $X=DZ+E$\\
\hline
LRRi & $\left\|Z\right\|_*+\lambda\left\|E\right\|_1+\frac{\gamma}{2}\left\|E-E_0\right\|_F^2$ & $X=DZ+E$\\
\hline
GSR & $\left\|Z^T\right\|_{2,1}+\lambda\left\|E\right\|_1$ & $X=DZ+E$\\
\hline
GSRi & $\left\|Z^T\right\|_{2,1}+\lambda\left\|E\right\|_1+\frac{\gamma}{2}\left\|E-E_0\right\|_F^2$ & $X=DZ+E$\\
\hline
\end{tabular}
\end{table}

\section{Informed Group-Sparse Representation}
\label{sec:gsri}

In jazz and popular music, it is well known that a few chord symbols are enough to compactly represent the harmonic structure of a piece. To motivate our new representation, let us begin with a simple chord sequence C-G-F-G-C for the instrumental part (see \cite{Coker64} for chord notations). If we have a learned dictionary with the C, Dm, Em, F, G, Am, and Bm$\flat 5$ chords, then the C, F, and G chords can be represented as:
\begin{align}
\label{eq:8}
C=\begin{pmatrix}
1 & 0 & 0 & 0 & 0 & 0 & 0
\end{pmatrix}^T,\\
F=\begin{pmatrix}
0 & 0 & 0 & 1 & 0 & 0 & 0
\end{pmatrix}^T,\\
G=\begin{pmatrix}
0 & 0 & 0 & 0 & 1 & 0 & 0
\end{pmatrix}^T,
\end{align}
and the time-atom representation of C-G-F-G-C becomes:
\begin{equation}
\label{eq:9}
Z=\begin{pmatrix}
1 & 0 & 0 & 0 & 1\\
0 & 0 & 0 & 0 & 0\\
0 & 0 & 0 & 0 & 0\\
0 & 0 & 1 & 0 & 0\\
0 & 1 & 0 & 1 & 0\\
0 & 0 & 0 & 0 & 0\\
0 & 0 & 0 & 0 & 0
\end{pmatrix}.
\end{equation}
One observation is that there are many empty rows in this representation (because not all the chords in the dictionary are used in the given sequence). So, a promising strategy for the inverse problem is to encourage row sparsity given an instrumental dictionary. Together with the idea of informed separation incorporating vocal annotations \eqref{eq:6}, we arrive at the following formulation:
\begin{align}
\label{eq:10}
\min_{Z,E}\left\|Z^T\right\|_{2,1}+\lambda\left\|E\right\|_1+\frac{\gamma}{2}\left\|E-E_0\right\|_F^2\\\nonumber
\text{s.t.}~X=DZ+E,
\end{align}
where $X$ is the input spectrogram, $D$ is the instrumental dictionary, $E_0$ denotes the vocal annotations, $DZ$ is the separated instrumentals, $E$ is the separated vocals, and $\left\|Z^T\right\|_{2,1}=\sum_i\sqrt{\sum_j Z_{ij}^2}$ denotes the sum of the $\ell_2$-norms of the rows of $Z$. As row sparsity is a kind of group sparsity \cite{Yuan06}, we call this the informed group-sparse representation (GSRi). In case where the vocal annotations are unavailable, we set $\gamma$ to zero, which we simply call the group-sparse representation (GSR). Our observation is further strengthened by the fact that group sparsity has been successfully applied to other audio processing models before \cite{Lefevre11,Hanlon16,Kowalski13}.

\subsection{Optimization}

The above formulation is not trivial to solve since the $\left\|\cdot\right\|_{2,1}$ and $\left\|\cdot\right\|_1$ norms are nonsmooth. Moreover, there is an additional equality constraint to be satisfied. In this case, the alternating direction method of multipliers (ADMM) \cite{Ma16} can be applied. ADMM works by first rewriting the constraint(s) into an augmented Lagrange function, then updating each variable in an alternating fashion until convergence. Although the convergence of ADMM has not been fully proven, it often converges in practice (cf.~\cite{Yang13}). Thus, to solve \eqref{eq:10}, we first introduce two auxiliary variables $J$ and $B$ for the alternating updates and rewrite the optimization problem as follows:
\begin{align}
\label{eq:11}
\min_{Z,J,E,B}\left\|J^T\right\|_{2,1}+\lambda\left\|B\right\|_1+\frac{\gamma}{2}\left\|E-E_0\right \|_F^2\\\nonumber
\text{s.t.}~X=DZ+E,~Z=J,~E=B.
\end{align}
The unconstrained augmented Lagrangian $\mathcal{L}$ is given by:
\begin{align}
\label{eq:12}
\mathcal{L}= & \left\|J^T\right\|_{2,1}+\lambda\left\|B\right\|_1+\frac{\gamma}{2}\left\|E-E_0\right\|_F^2\\\nonumber
& +\left\langle Y_1,X-DZ-E\right\rangle+\left\langle Y_2,Z-J\right\rangle+\left\langle Y_3,E-B\right\rangle\\\nonumber
& +\frac{\mu}{2}\left(\left\|X-DZ-E\right\|_F^2+\left\|Z-J\right\|_F^2+\left\|E-B\right\|_F^2\right)
\end{align}
where $Y_1$, $Y_2$, and $Y_3$ are the Lagrange multipliers. We then iteratively update the solutions for $J$, $Z$, $B$, and $E$.

\subsubsection{Updating J}

By minimizing \eqref{eq:12} with respect to $J$, we get a closed-form solution by groupwise soft thresholding \cite{Yuan06}:
\begin{align}
\label{eq:13}
J & =\arg\min_J\left\|J^T\right\|_{2,1}+\frac{\mu}{2}\left\|J-\left(Z+\mu^{-1}Y_2\right)\right\|_F^2\\\nonumber
& =\left(\left(1-\frac{\lambda}{\left\|\left(Z+\mu^{-1}Y_2\right)_i\right\|}\right)_+\left(Z+\mu^{-1}Y_2\right)_i\right)_{i=1}^k,
\end{align}
where $k$ is the number of rows of $J$, $A_i$ denotes the $i$-th row of $A$, $\left(A_i\right)_{i=1}^k=\left(A_1^T\ldots A_k^T\right)^T$, and $\left(a\right)_+=\max(0,a)$.

\subsubsection{Updating Z}

By differentiating \eqref{eq:12} with respect to $Z$ and setting $\partial\mathcal{L}/\partial Z=0$, we have:
\begin{equation}
\label{eq:14}
Z=\left(I+D^TD\right )^{-1}\left(D^T\left(X-E\right)+J+\mu^{-1}\left(D^TY_1-Y_2\right)\right).
\end{equation}
The solutions for $B$ and $E$ can be obtained analogously. Finally, we update the Lagrange multipliers as in \cite{Ma16}.\footnote{In order to encourage reproducible research, all the code for our paper is made available at \url{http://mac.citi.sinica.edu.tw/ikala/code.html}.} The algorithm is linear time and it does not rely on the SVD, for it uses the $\ell_{2,1}$-norm instead of the trace norm. Given input $X\in\mathbb{R}^{m\times n}$ and $D\in\mathbb{R}^{m\times k}$, if we assume $k\ll n$ and $m\ll n$, and further assume the number of iterations to be small, then the running time of our algorithm is $O{\left(kn\left(k+m\right)\right)}\approx O{\left(n\right)}$. Following \cite{Lin09}, we use $\mu\gets\rho\mu$ at each iteration to obtain a nondecreasing sequence of $\mu$. For pure ADMM, we should fix $\rho=1$; however, we can use higher values for faster convergence in practice. The faster variant ($\rho>1$) is known as the inexact augmented Lagrangian method (IALM) \cite{Lin09}.

\subsection{Relation to Low-Rank Representation}

In Section~\ref{sec:intro}, we have seen that LRR is equivalent to RPCA when $D=I$. There is a similar relation between LRR and GSR. We can factorize the matrix $Z$ as follows (cf.~\cite{Grave11}):
\begin{equation}
\label{eq:17}
Z=I_k\ diag{\left(\left\|Z_1\right\|,\ldots,\left\|Z_k\right\|\right)}\begin{pmatrix}
Z_1/\left\|Z_1\right\|\\
\vdots\\
Z_k/\left\|Z_k\right\|
\end{pmatrix}.
\end{equation}
If $Z$ has orthogonal rows, then the above is also a valid SVD, since $I_k$ is orthonormal and the normalization above makes the rightmost term orthonormal too. As a consequence, we have:
\begin{equation}
\label{eq:18}
\left\|Z\right\|_*=\sum_{i=1}^k\left\|Z_i\right\|=\left\|Z^T\right\|_{2,1}.
\end{equation}
Given this condition, the equivalence between LRR and GSR can be easily established.

\begin{figure}
\centering
\includegraphics[width=\columnwidth]{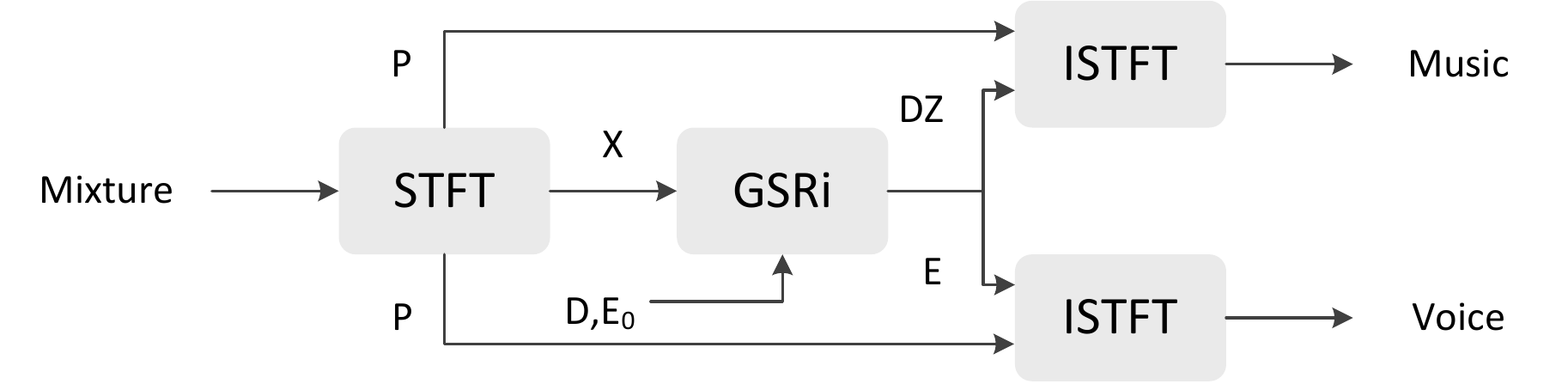}
\caption{Block diagram of our SVS system (see text for the variable definitions).}
\label{fig:3}
\end{figure}

\section{Experimental Results}
\label{sec:exp}

To evaluate the performance of GSRi, we use a source separation competition dataset called the iKala dataset \cite{Chan15}. This dataset contains $252~30$-second mono clips with human-labeled vocal pitch contours. The instrumentals and vocals are mixed at $0$ dB signal-to-noise ratio. We randomly select $44$ songs as the training set, leaving $208$ songs for the test set. The songs are downsampled from $44\,100$ Hz to $22\,050$ Hz to reduce memory usage, then a short-time Fourier transform (STFT) with a $1\,411$-point Hann window with $75$\% overlap is used to obtain the spectrograms \cite{Chan15}. The magnitude spectrogram $X$ is fed into GSRi and the separated components are reconstructed via inverse STFT using the original phase $P$. To get the vocal annotations, we first transform the human-labeled vocal pitch contours into a time-frequency binary mask. The authors of \cite{Ikemiya15} have proposed a harmonic mask similar to that of \cite{Virtanen08}, where it passes only integral multiples of the vocal fundamental frequencies (cf.~\cite{Ryynanen08,Durrieu11}):
\begin{equation}
\label{eq:19}
M(f,t)=\begin{cases}
1, & \text{if}~\left|f-nF_0(t)\right|<\frac{w}{2},\quad\exists n\in\mathbb{N}^+,\\
0, & \text{otherwise}.\\
\end{cases}
\end{equation}
Here $F_0(t)$ is the vocal fundamental frequency at time $t$, $n$ is the order of the harmonic, and $w$ is the width of the mask, which we set to $w=80$ Hz as in \cite{Ikemiya15}. Then we simply define the vocal annotations as $E_0=X\circ M$, where $\circ$ denotes the Hadamard product. Our experimental setup is shown in Fig.~\ref{fig:3}.

\subsection{Algorithms}

The algorithms to be compared are summarized in Table~\ref{tab:1}. All these methods (except RPCA and RPCAi) require prior training. For completeness, we further propose the informed LRR (LRRi) by simply replacing the $\ell_{2,1}$-norm by the trace norm. The resulting subproblem $J=\arg\min_J\left\|J\right\|_*+\frac{\mu}{2}\left\|J-\left(Z+\mu^{-1}Y_2\right)\right\|_F^2$ can be solved by singular value thresholding \cite{Cai10,Liu13}. The convergence criteria is $\left\|X-A-E\right\|_F/\left\|X\right\|_F<10^{-5}$, with $A=DZ$ if applicable. For $X\in\mathbb{R}^{m\times n}$, $\lambda$ is set to $1/\sqrt{\max(m,n)}$, and $\gamma$ is set to $2/\sqrt{\max(m,n)}$, following a grid search on the training set. All six algorithms are implemented from scratch using IALM with the same $\mu=10^{-3}$ and $\rho=1.2$ (see Section~\ref{sec:gsri}) to ensure fair timing comparisons.

\subsection{Dictionary}

We use non-negative sparse coding (NNSC) in the SPAMS toolbox \cite{Mairal09,Mairal10} to train our instrumental dictionary.\footnote{We have tried removing the non-negative constraints on $D$ and $\alpha$ but this does not change the results significantly for both datasets in this section.} Given $n$ input frames $x_i\in\mathbb{R}^m$, NNSC \cite{Hoyer02} learns a dictionary $D$ by solving the following joint optimization problem:
\begin{equation}
\label{eq:20}
\min_{D\geq 0,\alpha}\sum_{i=1}^n\frac{1}{2}\left\|x_i-D\alpha_i\right\|_2^2+\lambda\left\|\alpha_i\right\|_1~\text{s.t.}~\forall i,~\alpha_i\geq 0,
\end{equation}
where $\|\cdot\|_2$ denotes the Euclidean norm and $\lambda$ is a regularization parameter, which we set to $1/\sqrt{m}$ as in \cite{Mairal09}. The input frames are extracted from the training set after STFT. Following \cite{Yang13}, we define the dictionary size to be $100$ atoms.

\subsection{Evaluation}

For both the instrumentals and the vocals, separation performance is measured by BSS Eval toolbox version 3.0\footnote{\url{http://bass-db.gforge.inria.fr/}} in terms of source-to-distortion ratio (SDR), source-to-interferences ratio (SIR), and sources-to-artifacts ratio (SAR) \cite{Vincent06}, with higher values indicate better separation. We also compute the normalized SDR (NSDR) which is the improvement in SDR using the initial mixture as baseline \cite{Hsu10}. We then report the average result, denoted by the G prefix, for the test set. The most important measure is GNSDR as it measures the overall performance improvement. In addition, we also report the total running time of each algorithm on an IBM System x3650 M4 (two Intel E5-2697 v2 CPUs at $2.70$ GHz) with $384$ GB RAM.

\subsection{Results}

We can make several observations from the results and running times shown in Table~\ref{tab:2}. First, the informed algorithms (RPCAi, LRRi, GSRi) clearly outperforms their uninformed counterparts, ascertaining the usefulness of informed separation. Second, the performance of GSRi and LRRi are comparable, with GSRi performing slightly better in the music accompaniment part and LRRi performing slightly better in the vocal part. Third, the advantage of the learned dictionary can be shown by the superiority of GSR and LRR families over the RPCA family in the music accompaniment part. This means that the dictionary has successfully learned relevant information so that it performs better than plain sinusoids. Fourth, the running time of the GSR family is the fastest, showing the speed improvement by removing SVDs. Fifth, while the informed versions of RPCA and LRR are slower than the uninformed ones, this is not the case for GSR, for GSRi iterates much faster than LRRi. This makes GSRi more attractive than the alternatives. Finally, we remark that LRR is faster than RPCA because the SVD is applied to $Z$, which is much smaller than $A$.

\begin{table}
\caption{Results for voice (E) and music (A or DZ), in dB. The running time of each method, in hh:mm:ss, is also shown.}
\label{tab:2}
\centering
\begin{tabular}{|c|c|c|c|c|c|c|}
\hline
\multicolumn{2}{|l|}{} & GNSDR & GSDR & GSIR & GSAR & Runtime\\
\hline\hline
\multirow{2}{*}{RPCA} & E & 2.41 & 6.21 & 8.14 & 12.53 & \multirow{2}{*}{02:24:54}\\
& A & 4.48 & 0.76 & 3.23 & 7.00 &\\
\hline
\multirow{2}{*}{RPCAi} & E & \textbf{7.93} & \textbf{11.74} & \textbf{17.82} & 13.31 & \multirow{2}{*}{03:19:37}\\
& A & 10.89 & 7.17 & 13.31 & 9.00 &\\
\hline
\multirow{2}{*}{LRR} & E & 3.93 & 7.73 & 11.41 & 11.17 & \multirow{2}{*}{00:25:03}\\
& DZ & 5.42 & 1.70 & 3.40 & \textbf{9.63} &\\
\hline
\multirow{2}{*}{LRRi} & E & 7.75 & 11.55 & 16.92 & 13.38 & \multirow{2}{*}{00:30:12}\\
& DZ & 11.29 & 7.56 & 14.92 & 8.87 &\\
\hline
\multirow{2}{*}{GSR} & E & 2.50 & 6.30 & 7.36 & \textbf{14.80} & \multirow{2}{*}{00:13:16}\\
& DZ & 5.25 & 1.53 & 5.15 & 5.89 &\\
\hline
\multirow{2}{*}{GSRi} & E & 7.71 & 11.51 & 16.34 & 13.63 & \multirow{2}{*}{\textbf{00:13:15}}\\
& DZ & \textbf{11.31} & \textbf{7.59} & \textbf{15.19} & 8.82 &\\
\hline
\end{tabular}
\end{table}

\subsection{The Use of Multiple Dictionaries}

Suppose we concatenate $\kappa$ instrumental dictionaries together such that $D=\left(D_1\ldots D_\kappa\right)$. We can then apply GSRi directly to obtain the solution of the following:
\begin{align}
\label{eq:26}
\min_{Z,E}\left\|\left(Z_1^T\ldots Z_\kappa^T\right)\right\|_{2,1}+\lambda\left\|E\right\|_1+\frac{\gamma}{2}\left\|E-E_0\right\|_F^2\\\nonumber
\text{s.t.}~X=\left(D_1\ldots D_\kappa\right)\left(Z_1^T\ldots Z_\kappa^T\right)^T+E,
\end{align}
where $Z=\left(Z_1^T\ldots Z_\kappa^T\right)^T$. As $DZ=\sum_{i=1}^\kappa D_iZ_i$, both $DZ$ and $D_iZ_i$ can be interpreted as magnitude spectrograms: the former represents the instrumentals as a whole, while the latter represent the decomposed components associated with each dictionary. We call \eqref{eq:26} the informed multiple-group-sparse representation (MGSRi). To test whether multiple dictionaries is a feasible idea, we use the DSD100\footnote{\url{http://liutkus.net/DSD100.zip}} dataset which contains $50$ songs for training and $50$ for testing. To fit the SVS theme, we restrict ourselves to the pop/singer-songwriter subset.\footnote{\url{http://www.cambridge-mt.com/ms-mtk.htm}} Each song in DSD100 contains four sources (bass, drums, other, and vocals), so we can perform SVS by either:
\begin{itemize}
\item \emph{GSRi:} We mix bass, drums, and other equally into a single instrumental source. Then we learn the instrumental dictionary from this combined source and apply GSRi.
\item \emph{MGSRi:} We train $\kappa=3$ dictionaries for bass, drums, and other, respectively. After solving for $Z$ and $E$, the instrumentals as a whole ($DZ$) is used for comparison.
\end{itemize}
To reduce computations, we downmix to mono and downsample from $44\,100$ Hz to $22\,050$ Hz. As DSD100 does not have pitch contour labels, we create them by running MELODIA \cite{Salamon12} on the ground truth vocals. To eliminate the possible effect of dictionary size, we train the GSRi dictionary with $300$ atoms. Here we choose $w=60$ Hz and $\lambda=\gamma=1/\sqrt{\max(m,n)}$ for each $X\in\mathbb{R}^{m\times n}$, after a grid search on the training set. From the results in Table~\ref{tab:4}, we can conclude that MGSRi is not inferior to GSRi in SVS, with the biggest advantage that it can separate all the instrumental components (as in SiSEC MUS\footnote{\url{http://sisec.inria.fr/home/2016-professionally-produced-music-recordings/}}) while GSRi cannot.

\begin{table}
\caption{DSD100 (pop) results for voice (E) and music (DZ), in dB.}
\label{tab:4}
\centering
\begin{tabular}{|c|c|c|c|c|c|c|}
\hline
\multicolumn{2}{|l|}{} & GNSDR & GSDR & GSIR & GSAR\\
\hline\hline
\multirow{2}{*}{GSRi} & E & 8.08 & 4.64 & 11.24 & 6.11\\
& DZ & 5.49 & 8.94 & 13.36 & 11.30\\
\hline
\multirow{2}{*}{MGSRi} & E & 8.03 & 4.59 & 10.62 & 6.29\\
& DZ & 5.56 & 9.01 & 13.62 & 11.24\\
\hline
\end{tabular}
\end{table}

\section{Conclusion}
\label{sec:conc}

In this paper, we have proposed a novel GSRi method for SVS which incorporates both an instrumental dictionary and vocal annotations to inform the source separation process. Experimental results have shown that GSRi achieves the best performance in terms of instrumental GNSDR, GSDR, GSIR, and running time, making GSRi the best candidate for de-soloing applications. We have also successfully extended GSRi to the multiple-dictionary case. In conclusion, our experiments have shown that group sparsity achieves comparable results to low-rankness in a dictionary, but in a more efficient manner.

\bibliographystyle{IEEEtran}
\bibliography{chan17spl}

\begin{thebibliography}{10}
\providecommand{\url}[1]{#1}
\csname url@samestyle\endcsname
\providecommand{\newblock}{\relax}
\providecommand{\bibinfo}[2]{#2}
\providecommand{\BIBentrySTDinterwordspacing}{\spaceskip=0pt\relax}
\providecommand{\BIBentryALTinterwordstretchfactor}{4}
\providecommand{\BIBentryALTinterwordspacing}{\spaceskip=\fontdimen2\font plus
\BIBentryALTinterwordstretchfactor\fontdimen3\font minus
  \fontdimen4\font\relax}
\providecommand{\BIBforeignlanguage}[2]{{%
\expandafter\ifx\csname l@#1\endcsname\relax
\typeout{** WARNING: IEEEtran.bst: No hyphenation pattern has been}%
\typeout{** loaded for the language `#1'. Using the pattern for}%
\typeout{** the default language instead.}%
\else
\language=\csname l@#1\endcsname
\fi
#2}}
\providecommand{\BIBdecl}{\relax}
\BIBdecl

\bibitem{Comon10}
P.~Comon and C.~Jutten, \emph{{Handbook of Blind Source Separation}}.\hskip 1em
  plus 0.5em minus 0.4em\relax Oxford: Academic Press, 2010.

\bibitem{Tikhonov95}
A.~N. Tikhonov, A.~V. Goncharsky, V.~V. Stepanov, and A.~G. Yagola,
  \emph{{Numerical Methods for the Solution of Ill-Posed Problems}}.\hskip 1em
  plus 0.5em minus 0.4em\relax Dordrecht: Springer, 1995.

\bibitem{Tarantola05}
A.~Tarantola, \emph{{Inverse Problem Theory and Methods for Model Parameter
  Estimation}}, 2nd~ed.\hskip 1em plus 0.5em minus 0.4em\relax Philadelphia,
  PA: SIAM, 2005.

\bibitem{Vembu05}
S.~Vembu and S.~Baumann, ``{Separation of vocals from polyphonic audio
  recordings},'' in \emph{{Proc. Int. Soc. Music Inform. Retrieval Conf.}},
  2005, pp. 337--344.

\bibitem{Ryynanen08}
M.~Ryyn{\"a}nen, T.~Virtanen, J.~Paulus, and A.~Klapuri, ``{Accompaniment
  separation and karaoke application based on automatic melody
  transcription},'' in \emph{{Proc. IEEE Int. Conf. Multimedia and Expo}},
  2008, pp. 1417--1420.

\bibitem{Huang12}
P.-S. Huang, S.~D. Chen, P.~Smaragdis, and M.~Hasegawa-Johnson,
  ``{Singing-voice separation from monaural recordings using robust principal
  component analysis},'' in \emph{{Proc. IEEE Int. Conf. Acoust., Speech and
  Signal Process.}}, 2012, pp. 57--60.

\bibitem{Sprechmann12}
P.~Sprechmann, A.~Bronstein, and G.~Sapiro, ``{Real-time online singing voice
  separation from monaural recordings using robust low-rank modeling},'' in
  \emph{{Proc. Int. Soc. Music Inform. Retrieval Conf.}}, 2012, pp. 67--72.

\bibitem{Tsai12}
W.-H. Tsai and H.-C. Lee, ``{Automatic evaluation of karaoke singing based on
  pitch, volume, and rhythm features},'' \emph{{IEEE Trans. Audio, Speech,
  Language Process.}}, vol.~20, no.~4, pp. 1233--1243, 2012.

\bibitem{Nakano14}
T.~Nakano, K.~Yoshii, and M.~Goto, ``{Vocal timbre analysis using latent
  Dirichlet allocation and cross-gender vocal timbre similarity},'' in
  \emph{{Proc. IEEE Int. Conf. Acoust., Speech and Signal Process.}}, 2014, pp.
  5202--5206.

\bibitem{Vincent14}
E.~Vincent, N.~Bertin, R.~Gribonval, and F.~Bimbot, ``{From blind to guided
  audio source separation: How models and side information can improve the
  separation of sound},'' \emph{{IEEE Signal Process. Mag.}}, vol.~31, no.~3,
  pp. 107--115, 2014.

\bibitem{Cappe96}
O.~Capp{\'e} and E.~Moulines, ``{Regularization techniques for discrete
  cepstrum estimation},'' \emph{{IEEE Signal Process. Lett.}}, vol.~3, no.~4,
  pp. 100--102, 1996.

\bibitem{Kowalski08}
M.~Kowalski, E.~Vincent, and R.~Gribonval, ``{Under-determined source
  separation via mixed-norm regularized minimization},'' in \emph{{Proc.
  European Signal Process. Conf.}}, 2008, pp. 1--5.

\bibitem{Tibshirani96}
R.~Tibshirani, ``{Regression shrinkage and selection via the lasso},''
  \emph{{J. Roy. Stat. Soc. B}}, vol.~58, no.~1, pp. 267--288, 1996.

\bibitem{Foldiak90}
P.~F{\"o}ldi{\'a}k, ``{Forming sparse representations by local anti-Hebbian
  learning},'' \emph{{Biological Cybern.}}, vol.~64, pp. 165--170, 1990.

\bibitem{Olshausen97}
B.~A. Olshausen and D.~J. Field, ``{Sparse coding with an overcomplete basis
  set: A strategy employed by V1?}'' \emph{{Vision Research}}, vol.~37, no.~23,
  pp. 3311--3325, 1997.

\bibitem{Virtanen03}
T.~Virtanen, ``{Sound source separation using sparse coding with temporal
  continuity objective},'' in \emph{{Proc. Int. Comput. Music Conf.}}, 2003,
  pp. 231--234.

\bibitem{Plumbley10}
M.~D. Plumbley, T.~Blumensath, L.~Daudet, R.~Gribonval, and M.~E. Davies,
  ``{Sparse representations in audio and music: From coding to source
  separation},'' \emph{{Proc. IEEE}}, vol.~98, no.~6, pp. 995--1005, 2010.

\bibitem{Fazel01}
M.~Fazel, H.~Hindi, and S.~P. Boyd, ``{A rank minimization heuristic with
  application to minimum order system approximation},'' in \emph{{Proc. Amer.
  Control Conf.}}, 2001, pp. 4734--4739.

\bibitem{Srebro05}
N.~Srebro, J.~Rennie, and T.~S. Jaakkola, ``{Maximum-margin matrix
  factorization},'' in \emph{{Advances in Neural Information Processing Systems
  17}}, 2005, pp. 1329--1336.

\bibitem{Yang13}
Y.-H. Yang, ``{Low-rank representation of both singing voice and music
  accompaniment via learned dictionaries},'' in \emph{{Proc. Int. Soc. Music
  Inform. Retrieval Conf.}}, 2013, pp. 427--432.

\bibitem{Ikemiya15}
Y.~Ikemiya, K.~Yoshii, and K.~Itoyama, ``{Singing voice analysis and editing
  based on mutually dependent F0 estimation and source separation},'' in
  \emph{{Proc. IEEE Int. Conf. Acoust., Speech and Signal Process.}}, 2015, pp.
  574--578.

\bibitem{Chan15}
T.-S. Chan, T.-C. Yeh, Z.-C. Fan, H.-W. Chen, L.~Su, Y.-H. Yang, and R.~Jang,
  ``{Vocal activity informed singing voice separation with the iKala
  dataset},'' in \emph{{Proc. IEEE Int. Conf. Acoust., Speech and Signal
  Process.}}, 2015, pp. 718--722.

\bibitem{Bryan13}
N.~J. Bryan, G.~J. Mysore, and G.~Wang, ``{Source separation of polyphonic
  music with interactive user-feedback on a piano roll display},'' in
  \emph{{Proc. Int. Soc. Music Inform. Retrieval Conf.}}, 2013, pp. 119--124.

\bibitem{Liutkus13}
A.~Liutkus, J.-L. Durrieu, L.~Daudet, and G.~Richard, ``{An overview of
  informed audio source separation},'' in \emph{{Proc. Int. Workshop Image
  Anal. Multimedia Interactive Services}}, 2013, pp. 1--4.

\bibitem{Ewert14}
S.~Ewert, B.~Pardo, M.~Muller, and M.~D. Plumbley, ``{Score-informed source
  separation for musical audio recordings: An overview},'' \emph{{IEEE Signal
  Process. Mag.}}, vol.~31, no.~3, pp. 116--124, 2014.

\bibitem{Chen13}
\BIBentryALTinterwordspacing
Z.~Chen, P.-S. Huang, and Y.-H. Yang, ``{Spoken lyrics informed singing voice
  separation},'' in \emph{{Proc. HAMR}}, 2013. [Online]. Available:
  \url{http://labrosa.ee.columbia.edu/hamr2013/proceedings/doku.php/singing_separation}
\BIBentrySTDinterwordspacing

\bibitem{Lefevre14}
A.~Lef{\'e}vre, F.~Glineur, and P.-A. Absil, ``{A convex formulation for
  informed source separation in the single channel setting},''
  \emph{{Neurocomputing}}, vol. 141, pp. 26--36, 2014.

\bibitem{Jeong15}
I.~Y. Jeong and K.~Lee, ``{Informed source separation from monaural music with
  limited binary time-frequency annotation},'' in \emph{{Proc. IEEE Int. Conf.
  Acoust., Speech and Signal Process.}}, 2015, pp. 489--493.

\bibitem{Lin09}
Z.~Lin, M.~Chen, L.~Wu, and Y.~Ma, ``{The augmented Lagrange multiplier method
  for exact recovery of corrupted low-rank matrices},'' Tech. Rep.
  UILU-ENG-09-2215, 2009.

\bibitem{Candes11}
E.~J. Cand{\`e}s, X.~Li, Y.~Ma, and J.~Wright, ``{Robust principal component
  analysis?}'' \emph{{J. ACM}}, vol.~58, no.~3, pp. 1--37, 2011.

\bibitem{Rafii13}
Z.~Rafii and B.~Pardo, ``{REpeating Pattern Extraction Technique (REPET): A
  simple method for music/voice separation},'' \emph{{IEEE Trans. Audio,
  Speech, Language Process.}}, vol.~21, no.~2, pp. 73--84, 2013.

\bibitem{Su13}
L.~Su and Y.-H. Yang, ``{Sparse modeling for artist identification: Exploiting
  phase information and vocal separation},'' in \emph{{Proc. Int. Soc. Music
  Inform. Retrieval Conf.}}, 2013, pp. 349--354.

\bibitem{Yang12}
Y.-H. Yang, ``{On sparse and low-rank matrix decomposition for singing voice
  separation},'' in \emph{{Proc. ACM Multimedia}}, 2012, pp. 757--760.

\bibitem{Liu13}
G.~Liu, Z.~Lin, S.~Yan, J.~Sun, Y.~Yu, and Y.~Ma, ``{Robust recovery of
  subspace structures by low-rank representation},'' \emph{{IEEE Trans. Pattern
  Anal. Mach. Intell.}}, vol.~35, no.~1, pp. 171--184, 2013.

\bibitem{Hsu10}
C.-L. Hsu and J.-S.~R. Jang, ``{On the improvement of singing voice separation
  for monaural recordings using the MIR-1K dataset},'' \emph{{IEEE Trans.
  Audio, Speech, Language Process.}}, vol.~18, no.~2, pp. 310--319, 2010.

\bibitem{Salamon12}
J.~Salamon and E.~G{\'o}mez, ``{Melody extraction from polyphonic music signals
  using pitch contour characteristics},'' \emph{{IEEE Trans. Audio, Speech,
  Language Process.}}, vol.~20, no.~6, pp. 1759--1770, 2012.

\bibitem{Coker64}
J.~Coker, \emph{{Improvising Jazz}}.\hskip 1em plus 0.5em minus 0.4em\relax
  Englewood Cliffs, NJ: Prentice-Hall, 1964.

\bibitem{Yuan06}
M.~Yuan and Y.~Lin, ``{Model selection and estimation in regression with
  grouped variables},'' \emph{{J. Roy. Stat. Soc. B}}, vol.~68, no.~1, pp.
  49--67, 2006.

\bibitem{Lefevre11}
A.~Lef{\`e}vre, F.~Bach, and C.~F{\'e}votte, ``{Itakura-Saito nonnegative
  matrix factorization with group sparsity},'' in \emph{{Proc. IEEE Int. Conf.
  Acoust., Speech and Signal Process.}}, 2011, pp. 21--24.

\bibitem{Hanlon16}
K.~O'Hanlon, H.~Nagano, N.~Keriven, and M.~D. Plumbley, ``{Non-negative group
  sparsity with subspace note modelling for polyphonic transcription},''
  \emph{{IEEE/ACM Trans. Audio, Speech, Language Process.}}, vol.~24, no.~3,
  pp. 530--542, 2016.

\bibitem{Kowalski13}
M.~Kowalski, K.~Siedenburg, and M.~D{\"o}rfler, ``{Social sparsity!
  Neighborhood systems enrich structured shrinkage operators},'' \emph{{IEEE
  Trans. Signal Process.}}, vol.~61, no.~10, pp. 2498--2511, 2013.

\bibitem{Ma16}
S.~Ma, ``{Alternating proximal gradient method for convex minimization},''
  \emph{{J. Sci. Comput.}}, vol.~68, no.~2, pp. 546--572, 2016.

\bibitem{Grave11}
{\'E}.~Grave, G.~Obozinski, and F.~Bach, ``{Trace Lasso: A trace norm
  regularization for correlated designs},'' in \emph{{Advances in Neural
  Information Processing Systems 24}}, 2011, pp. 2187--2195.

\bibitem{Virtanen08}
T.~Virtanen, A.~Mesaros, and M.~Ryyn{\"a}nen, ``{Combining pitch-based
  inference and non-negative spectrogram factorization in separating vocals
  from polyphonic music},'' in \emph{{Proc. ISCA Tutorial and Research Workshop
  on Statistical and Perceptual Audition}}, 2008, pp. 17--20.

\bibitem{Durrieu11}
J.~L. Durrieu, B.~David, and G.~Richard, ``{A musically motivated mid-level
  representation for pitch estimation and musical audio source separation},''
  \emph{{IEEE J. Sel. Topics Signal Process.}}, vol.~5, no.~6, pp. 1180--1191,
  2011.

\bibitem{Cai10}
J.~Cai, E.~J. Cand{\`e}s, and Z.~Shen, ``{A singular value thresholding
  algorithm for matrix completion},'' \emph{{SIAM J. Optimization}}, vol.~20,
  no.~4, pp. 1956--1982, 2010.

\bibitem{Mairal09}
J.~Mairal, F.~Bach, J.~Ponce, and G.~Sapiro, ``{Online dictionary learning for
  sparse coding},'' in \emph{{Proc. Int. Conf. Mach. Learning}}, 2009, pp.
  689--696.

\bibitem{Mairal10}
------, ``{Online learning for matrix factorization and sparse coding},''
  \emph{{J. Mach. Learning Research}}, vol.~11, pp. 19--60, 2010.

\bibitem{Hoyer02}
P.~O. Hoyer, ``{Non-negative sparse coding},'' in \emph{{Proc. IEEE Workshop
  Neural Networks Signal Process.}}, 2002, pp. 557--565.

\bibitem{Vincent06}
E.~Vincent, R.~Gribonval, and C.~F{\'e}votte, ``{Performance measurement in
  blind audio source separation},'' \emph{{IEEE Trans. Audio, Speech, Language
  Process.}}, vol.~14, no.~4, pp. 1462--1469, 2006.

\end{thebibliography}

\end{document}